# Social Secret Sharing for Resource Management in Cloud

(Technical Report Based on Ref. 5)

**Sandeep.R.Narani**
**narani@siu.edu**
Department of Computer Sciences
Southern Illinois University Carbondale, IL, USA

**Abstract:** *We first explain the notion of secret sharing and also threshold schemes, which can be implemented with the Shamir's secret sharing. Subsequently, we review social secret sharing [6,10] and its trust function. In a secret sharing scheme, a secret is shared among a group of players who can later recover the secret. We review the construction of a social secret sharing scheme and its application for resource management in cloud, as explained in [5]. To clarify the social secret sharing scheme, we first review its trust function according to [9]. In this scheme, a secret is maintained by assigning a trust value to each player based on his behavior, i.e., availability*.

## 1. Secret Sharing

Secret Sharing Schemes were discovered independently by Shamir and Blakely [3, 18]. The motivation for Secret Sharing is secure key management. In some situations, there is usually one secret key that provides access to many important files. If such a key is lost (e.g., the person who knows the key becomes unavailable, or the computer which stores the key is destroyed), then all the important files become inaccessible. The basic idea in Secret Sharing [3, 18] is to divide the secret key into pieces and distribute the pieces to different persons so that certain subsets of the persons can get together to recover the key. Secret Sharing refers to a method for distributing a secret among a group of people or participants, each of whom is allocated a share of the secret. This secret can be reconstructed only when a sufficient number of possibly different types of share are combined together. Individual single shares are of no use on their own in construction of the secret.

Example, suppose you and your friend suddenly got a map, that you believe would lead to an island, full of treasure of wealth. You and your friend are very happy and would like to go home and imagine for the exciting journey to the great fortune. Now, who is going to keep the map? If suppose you and your friend do not really trust each other and are afraid that, if the other one has the map might just go alone and take everything. Now we need a scheme that could make sure that the map is share in a way so that no one would be left out in this trip. The best way to come out of the problem is to split the map into two pieces and make sure that both pieces are needed in order to find the island of the treasure. Now each of them has a piece of the map and is pretty sure that the other cannot reach the treasure without their help [1].

**Main Arena:**

• Secret: The document which is kept unknown for most of the members in the crew.
And generally this document consists of any secure keywords or encrypted data.

• Parties: The equipment required storing the key or the secret document, it might be computers or deposit box, or memory sticks, etc., the devices which are respectively used for storing purposes.

• Share: The piece of the key allocated to respective personality. The combination of all the shares will lead to the final product.

These three areas remain the important aspects of the Secret Sharing. Losing or misplacing of any of these, may lead to disastrous results.

## 1.1 Types of Secret Sharing

There are many Secret Sharing deals with schemes, which were in use in early days for securing data transmission between two or more shareholders.

One type of secret sharing scheme is *one dealer and 'n' players* which is the basic idea of Shamir's Secret Sharing Scheme [3]. The dealer gives a share of the secret to the players, but only when specific conditions are fulfilled, the players will be able to reconstruct the secret from their shares. The dealer accomplishes this by giving each player a share in such a way that any group of 't' for threshold values or more players can together reconstruct the secret but no group of fewer than 't' players can. Such a system is called a (t, n)-threshold scheme (or (n, t)-threshold scheme).

The other type of secret sharing scheme is *geometric in nature* and is the idea of Blakley's Secret Sharing Scheme [18] but it is not to the perfection as it deals with hyperplane. Nevertheless, this scheme can be modified to achieve perfect security.

In this project, I am trying to review on the notion of Social Secret Sharing [10] and its trust functionality along with, how this construction can be used in cloud computing to create a self-organizing environment. In fact, distributed secure systems using threshold secret sharing can be adjusted automatically based on the resource availability of the cloud providers. It is also been illustrated how social secret sharing can be applied in the distributed systems using cloud computing technology which shares the files and allows for quick transfer of the shares. Paper contributions on the cryptographic primitive can be used to create self -organizing protocol in the cloud. It also states that distributed system can be reconfigured automatically based on the resource availability of the cloud providers by using social secret sharing scheme [5].

## 2. Introduction to Secret Sharing Schemes

In modern cryptography, the security of data is fully dependent on the security of the keys used. As most of the ciphers are public knowledge, one can easily encrypt and decrypt any message if they know the key involved. For some highly confidential data, it's not always good to have a single person in control of the key and to secure of the data. This has lead to the need for Secret Sharing Schemes, which allow keys to be distributed among a group of people, with a pre-specified number of them needing to input their share in order, to access the key. Secret Sharing scheme is a cryptographic technique in the security area which allows confidential data to be split or shared among several storage providers. Individually, each provider will learn absolutely nothing about these data. Combined, a designated group of providers will be able to recover the data with the help of this scheme. Here are the mathematical techniques for constructing Secret Sharing Schemes followed by the applications of secret sharing schemes [8].

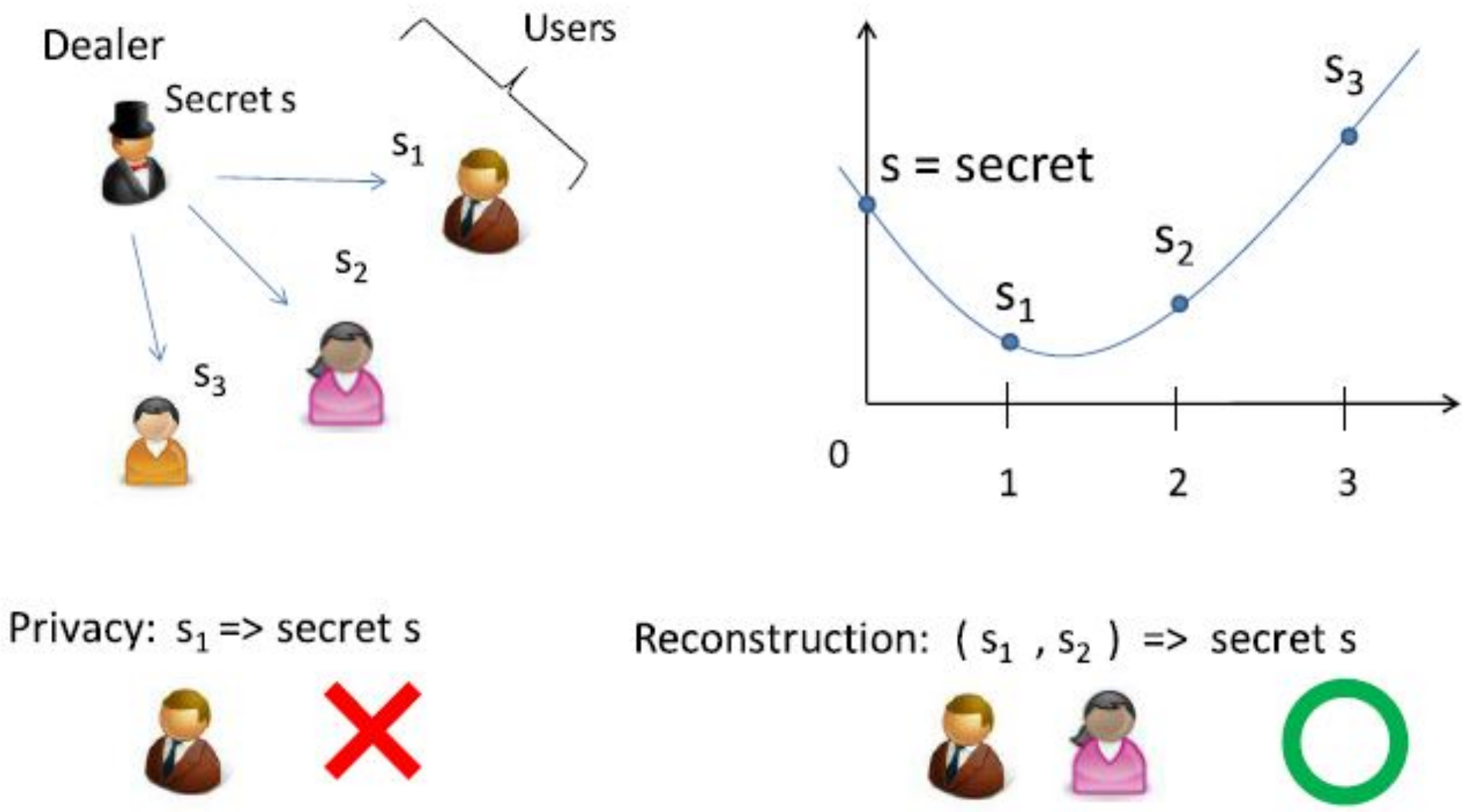

**Fig 1. Secret Sharing [8]**

## 2.1 Secret Sharing Schemes Properties

Basic Secret Sharing Schemes have two fundamental properties

1. Secrecy: Unauthorized subsets of participants should be prevented from learning the secret.
2. Correctness: Authorized subsets of participants should be able to recover the secret by pooling their shares.

And most of the other properties of the Secret Sharing Schemes are extensively studied by

- **Mathematicians** as objects of intrinsic interest in their own right.

- **Cryptographers** as important cryptographic primitives.
- **Security engineers** as techniques to employ in distributed security applications.

Consider a secret shared between friends in technical terminology, mostly Secret Sharing Schemes involves two hidden entities.

- The *dealer* is the entity normally responsible for:
  - generating system parameters
  - generating the secret
  - creating initial shares
  - sending initial shares to participants

- The *combiner* is the entity responsible for:
  - pooling shares
  - reconstructing the secret

## 3. Schemes

### 3.1 Shamir's Secret Sharing Scheme & implementation

Old basic Secret Sharing technique is Shamir's scheme which is based on polynomial interpolation & Lagrange interpolation formula given any k pairs $(x_1, y_1), \ldots, (x_k, y_k)$ with
$x_i = x_j$ for all $1 \le i < j \le k$, there is one and only one polynomial P(x) of degree $k - 1$ such that
$P(x_i) = y_i$, for all $1 \le i \le k$ [3].

• The secret 'S' is selected as the free coefficient in the random polynomial P of degree $k - 1$ over the field of the positive integers modulo a large prime.
• The shares $I_1, \ldots, I_n$ are chosen as $I_i = P(x_i)$, for all $1 \le i \le n$, where $x_1, \ldots, x_n$ are pair wise and bitwise distinct public values.
• Having the shares $\{I_{i|i} \in A\}$, for some group A with $|A| = k$, the secret can be obtained using Lagrange interpolation formula as

$$S = \sum_{i \in A} \left( \mathbf{I}_i \prod_{j \in A \setminus \{i\}} \cdot \frac{x_j}{x_j - x_j} \right)$$

## 3.2 Threshold Secret Sharing Schemes & Examples:

In the first Secret Sharing Schemes only the number of the participants in the reconstruction phase was important for recovering the secret. Such schemes have been referred to as threshold secret sharing schemes.

Let $n \geq 2$, $2 \leq k \leq n$. The access structure $A = \{A \in P(\{1, 2, . . . , n\}) \mid |A| \geq k\}$ will be referred to as the (k, n)-threshold access structure .We obtain $A_{min}= \{A \in P(\{1, 2, . . . , n\}) \mid |A| = k\}$, $Á= \{A \in P(\{1, 2, . . . , n\}) \mid |A| \leq k - 1\}$, and $Á_{max}= \{A \in P(\{1, 2, . . . , n\}) \mid |A| =k - 1\}$.

In this an Á secret sharing scheme will be shown as (k, n) threshold secret sharing scheme.

If $P(x) = a_{k-1}x^{k-1} + \cdot \cdot \cdot + a_1x^1 + a_0$, the secret can also be have the shares $I_{i1}, . . . , I_{ik}$ by solving the system of equations

$$\begin{cases} \mathbf{a_{k-1}x_{i1}^{k-1} + \cdot \cdot \cdot + a_1x_{i1}^1 + a_0 = I_{i1}} \\ . \\ . \\ \mathbf{a_{k-1}x_{ik}^{k-1} + \cdot \cdot \cdot + a_1x_{ik}^1 + a_0 = I_{ik}} \end{cases}$$

Where has k unknowns $(a_{k-1}, . . . , a_1, a_0)$ and it has a different solution because the determinant of as follow as

$$\begin{vmatrix} \mathbf{x_{i1}^{k-1} \cdot \cdot \cdot x_{i1}^1} & \mathbf{1} \\ \mathbf{x_{i2}^{k-1} \cdot \cdot \cdot x_{i2}^1} & \mathbf{1} \\ \mathbf{...} & \\ \mathbf{. . .} & \\ \mathbf{x_{ik}^{k-1} \cdot \cdot \cdot x_{ik}^1} & \mathbf{1} \end{vmatrix}$$

In the non-zero Vandermonde determinant. The point can be view, the polynomial P(x) can be chosen of degree at most $k - 1$.

With having only k−1 shares, the system of equations, in Shamir’s equation.

$$\begin{cases} \mathbf{a_{k-1}x_{i1}^{k-1} + \cdot \cdot \cdot + a_1x_{i1}^1 = I_{i1} - a_0} \\ . \\ . \\ \mathbf{a_{k-1}x_{ik}^{k-1} + \cdot \cdot \cdot + a_1x_{ik-1}^1 = I_{ik-1} - a_0} \end{cases}$$

With k−1 equations and k−1 unknowns ($a_{k-1}$,. ........., a1) has a different solution, for any $a_0$. Therefore, all possible values of the secret are equally likely.

The degree of the polynomial p(x) is known to be k-1 and it is remarkable or equivalent $a_{k-1} \neq 0$ then therefore scheme is not perfect. From the, any k − 1 users can determine an element $b_0$ which is not the secret, i.e., $b_0 \neq a0$, A polynomial $Q(x) = b_{k-2}x^{k-2} + \cdots + b_1x^1 + b_0$can be determined using Lagrange interpolation formula, such that $Q(x_{ij}) = I_{ij} = P(x_{ij})$, for all $1 \leq j \leq k-1$, leading to the system [4].

$$\begin{cases} \mathbf{a_{k-1}x_{i1}^{k-1} + (a_{k-2} - b_{k-2})x_{i1}^{k-2} + \ldots\ldots\ldots\ldots\ldots\ldots + (a_1 - b_1)x_{i1}^{1} + (a_0 - b_0) = 0} \\ \mathbf{a_{k-1}x_{ik-1}^{k-1} + (a_{k-2} - b_{k-2})x_{ik-1}^{k-2} + \ldots\ldots\ldots\ldots + (a_1 - b_1)x_{ik-1}^{1} + (a_0 - b_0) = 0} \end{cases}$$

Let us consider the contradiction, that $a_0 = b_0$, from the above function with k −1 equations and k − 1 unknowns $(a_{k-1}, \ldots\ldots, a_1)$ has in this case a different solution, namely
$a_{k-1} = 0, a_{k-2} = b_{k-2}, \ldots, a_1 = b_1$ that contradicts that $a_{k-1} \neq 0$. Therefore, any k − 1 users can determine an element $b_0$ which is not the secret and eventually, their uncertainty about the secret does not tally with the uncertainty of an outsider.

Shamir [3] has proposed choosing $x_i = i$, for all $1 \leq i \leq n$. In this case, the secret can be reconstructed as for any group A with |A| = k [3].

$$S = \sum_{i \in A} (\mathbf{Ii} \, . \prod_{j \in A \setminus \{i\}} . \frac{j}{j - i})$$

***Example 1***

Let n = 5 and k = 3. Let us consider the polynomial $P(x) = 2x^2 + 7x + 10$ over the field $Z_{11}$. The secret is S = 10 and the corresponding shares are the corresponding shares are

$P(1) = 2(1)^2 + 7(1) + 10$ => 19 mod 11 =8

$P(2) = 2(2)^2 + 7(2) + 10$ =>32 mod 11 =10

$P(3) = 2(3)^2 + 7(3) + 10$ =>49 mod 11 =5

$P(4) = 2(4)^2 + 7(4) + 10$ =>70 mod 11 =4

$P(5) = 2(5)^2 + 7(5) + 10$ =>95 mod 11 =7

Having the shares p (1), p (2), p (3) the secret can be reconstructed as

$= 8 . \frac{2}{2-1} . \frac{3}{3-1} + 10 \frac{1}{1-2} . \frac{3}{3-2} + 5 \frac{1}{1-3} . \frac{2}{2-3}$

=> (-1) mod 11 = 10

***Example.*2**

Shamir Secret Sharing with $p = 31$. Let the threshold be $t = 3$, and the secret be $7 \in Z/_{31Z}$. We choose elements at random $a_1 = 19$ and $a_2 = 21$ in $Z/_{31Z}$, and set $f(x) = 7 + 19x + 21x^2$. As the trusted party, we can now generate as many shares as we like,

(1, f(1)) = (1, 16)  (5, f(5)) = (5, 7)

(2, f(2)) = (2, 5)   (6, f(6)) = (6, 9)

(3, f(3)) = (3, 5)    (7, f(7)) = (7, 22)

(4, f(4)) = (4, 16) (8, f(8)) = (8, 15)

Which are distributed to the holders of the share recipients, and the original polynomial f(x) is destroyed.

The secret can be recovered from the first three shares (1, 16), (2, 5), (3, 5), we compute

$$f(0) = \frac{16 \cdot 2 \cdot 3}{(1-2)(1-3)} + \frac{5 \cdot 1 \cdot 3}{(2-1)(2-3)} + \frac{5 \cdot 1 \cdot 2}{(3-1)(3-2)}$$

$$\Rightarrow 3 \cdot 2^{-1} + 15 \cdot (-1) + 10 \cdot 2^{-1} = 17 - 15 + 5 = 7$$

By using different calculation for the shares (1, 16), (5,7), and (7, 22),

$$f(0) = \frac{16 \cdot 2 \cdot 3}{(1-2)(1-3)} + \frac{7 \cdot 1 \cdot 7}{(5-1)(5-7)} + \frac{22.1.5}{(7-1)(7-5)}$$

$$\Rightarrow 2 \cdot 24^{-1} + 18 \cdot (-8)^{-1} + 17 \cdot 12^{-1} = 13 + 21 + 4 = 7$$

Shamir himself has remarked, his scheme has some interesting features:

- In the Shamir's scheme is ideal when the size of every share doesn't exceed  the size of the secret .
- It is active, in sense that if the threshold 'k' is kept fixed, some existing secret can be removed or some new secret can be generated, without affecting the other secrets.

  McEliece and Sarwate [2] have remarked that Shamir's scheme is closely related to Reed-Solomon codes and that decoding algorithms for such codes can be used for generalizing Shamir's scheme.

Some Difference among the Secret Sharing Schemes:

- Verifiable Secret Sharing [7]
- Chinese Remainder Scheme (schemes based on the Chinese Remainder Theorem) [19]
- Secret Sharing Homomorphism (useful for many applications) [20]
- Visual Secret Sharing (secret and shares are images) [21]
- Black box Secret Sharing (schemes that are independent of the underlying group) [22]
- Anonymous Secret Sharing (identities of participants not required for reconstruction) [23]
- Ideal Secret Sharing (Shares equally to the size) [24]
- Weight Secret Sharing [32]

## 4. Adversary Model

"A system without an adversary definition cannot possibly be insecure; it can only be astonishing".

"… Astonishment is a much underrated security vice." (Principle of Least Astonishment)

From the above stated statements, it can be explained that any system without an adversary, hardly can be believed. To have a highly secured system, we also need to plan about the adversaries the system may face in the future. The adversary model consists of set of assumptions, explicit and implicit, which have been made with regard to the adversary in any given situation. While there is precedence for using such a definition of adversary modeling, it is not widely used in literature and there are some changing properties in this model which are related to the traditional secret sharing model.

• Trusted dealer: An adversary cannot corrupt the dealer, who is fully trusted.
• Polarized participants: Participants are either completely honest which follows the protocol.
or completely malicious and they have been captured by an adversary, who will attempt to subvert the protocol [17].

In all Secret Sharing Schemes, an adversary wishes to learn information about the secret, perhaps, through learning information about shares. However, with respect to the recoverability, the goals of different types of Secret Sharing Schemes vary. In the traditional model, the passive adversary can only engage in share capture and hence they can only try to prevent reconstruction of the secret through withholding shares. For each of the different types of Secret Sharing [3, 18], we shall begin, by identifying the main recoverability goals of an adversary.

## 4.1 Types of adversary Model

**Passive versus Active Adversary:** Every player has a curiosity to know the other players secret, but players feels they are honest to them. At the same time they have the curiosity to know the secret, which is an adversary. When players want to form the network or game, they will reconstruct an incorrect secret while trying to know the secret of others [5].
**Static versus Mobile Adversary:** The former refers to an adversary who corrupts the players ahead of time whereas the latter refers to an adversary who corrupts the players while the protocol is executing [5].
**Computational versus Unconditional Security:** The network protocols are much secured and relies on basic presumptions (the hardness of factoring or discrete logarithm), from the adversary has unlimited computational power [5].

## 5. Verifiable Secret Sharing Schemes (VSS)

What happens if in the middle of the procedure the players change their thinking, in terms of secrecy? Or change the content into a malicious data? Etc.,
To overcome the above problems VSS schemes [7] came into existences. In these schemes, shares are verifiable without revealing their secret and here we can also detect the malicious dealer causing changes to the content. In these schemes, every players can validate the compliance of the shares in both sharing and recovery phases. The authors [30, 31] provide the first unconditionally secure VSS when $t < (n/3)$ with a zero probability of error. In this scenario, each pair of the players is connected in the network with a secure private channel. To receive a higher threshold t < (n/2), the existence of both private channels and broadcast channels is required. This protocol t < (n/4) has an insignificant probability of error. To verify the constructions, verifiable schemes based on symmetric bivariate polynomials are in an unconditional secured format [5].

**Properties of VSS:**

- Do not assume a trusted dealer.
- Only authorize group of participants are given access to learn the secret.
- Honest participants want to recover secret even if adversary corrupts the dealer and some shares.
- Main recoverability goal of adversary is to prevent correct secret from being reconstructed.

VSS scheme have an additional algorithm, which allows participants to check and perform the functions.

**Consistency**: Upon, Verification the authorized group of participants, who accept their shares, will be able to reconstruct the same Secret.

**Correctness**: If dealer was honest then value is the genuine secret.

**Types of VSS scheme**

1. Interactive VSS: If verification involves participants exchanging messages between themselves. [7].
2. Non-interactive VSS: If Verification involves only participants exchanging messages with the dealer.
3. Publicly: Verifiable, if honest participants are assured of the validity of their own share and the shares of other participants.

The notion of proactive secret sharing (PSS) [25] is proposed where the shares of the players are updated without changing the secret. This can be done by adding the shares of a new polynomial 'g' with zero constant term to the shares of the original secret sharing polynomial f with constant term ξ. As a result, the new secret sharing polynomial will be $\hat{f}$ = f +g where $\hat{f}$ (0) = ξ. In other words, players frequently change the secret sharing polynomial to deal with a mobile adversary who can incrementally collect the shares of the players while the protocol is executing. To assign multiple shares rather than a single share to some players, Weighted Secret Sharing (WSS) [32] is introduced which is used to prioritize different players in a hierarchy structure.

It can be illustrate by using WSS how social secret sharing [10] can be applied in distributed secure systems by using cloud computing infrastructures. It can improve social secret sharing by proposing a new trust function. Here is a scenario in which the cryptographic primitive can be used to create a self-organizing protocol in the cloud. In fact, it has been shown that a distributed system can be reconfigured automatically based on the resource availability of the cloud providers. Subsequently a new trust function with social properties in order to improve the existing social secret sharing scheme is provided [5].

## 6. Social Secret Sharing

In this scheme [6, 10], each participant or player will receive a constant number of shares during the start. As time goes on, these players are assigned to the weights based on their behaviors in the scheme. As results, which have the order to each player receives a number of shares corresponding to his trust value. The weights of each participant are adjusted in such a way that cooperative players receive more shares compared to non-cooperative ones. It alters; new participate to join the scheme while corrupted players are disconnected immediately. The construction of a trust function is independent of our proposed scheme.

Therefore, we use the trust management approach proposed in [9], they defined six possible actions (i.e., encourage, give a chance, reward, penalize, take a chance, and discourage). These are the view held by the author views eventually, they apply monotonically increasing and decreasing functions in the case of cooperation and defection in order to compute players trust values. Now, the authors [10] can modify the major definitions required for our schemes.

In this study, I made findings that are easy to comprehend and apply as well. The number of shares assigned to each person depends on that person's level of work and the way he communicates with different people. It can be implied that the weight of this person is managed such that people who are participating along with him also get more shares when compared to that of the people who are not cooperating. Similarly, in our social life, we exchange or share our personal things and secrets only with the people whom we have trust on and vice versa.

***Definition 1:*** *Social Secret Sharing is a three-tuple denoted as (Sha, Tun, Rec) consisting of Secret Sharing, social tuning, and secret recovery respectively. The only difference compared to threshold Secret Sharing is the second stage, in which the weight of each player $P_i$ is adjusted according to his reputation [5].*

**A. Assumptions:**

We have made some assumptions on the properties which are necessary to build a Secret Sharing Scheme. Consider a secret ξ which is about to be shared:

1) In order to retrieve this secret ξ, the total weight of selected persons $P_i \in \Delta$ must be equal or greater than the threshold:

$$\sum_{P_i \in \Delta} w_i \geq t$$

Here Δ denotes the set of uncorrupted members participating in the plan. It can be further shown that this set is then classified divided into three sub-groups N - new, B - bad and G - good that denote the people who are new, non-cooperating, and cooperating people respectively.

2) Along with this the total weight of colluders $P_i \in \nabla$ should not be more than the threshold, where ∇ denotes the set of corrupted players:

$$\sum_{Pi \in \nabla} w_i < t$$

3) Then the weight of each person $P_i$ is limited by an attribute which is very much less than t, that is, $w_i \leq m \ll t$ for $1 \leq i \leq n$.

**B. Weight optimization:**

In this Social Secret Sharing Scheme [6], a selected person initially gets many numbers of shares from the sender. Eventually, the plan is modified based on the characteristics of the participants. The social tuning phase is further reviewed by following a set of steps. In order to make it simple, let's consider making the weights increment or decrement in a step-by-step procedure.

**Adjustment**: This is made on the basis of the available participants or the response time. Here the "reputation" and consequently the "weights" of all the players are adjusted.

**Enrollment:** In order to incline the weight of a participant who cooperates, these groups combine to create a new share on the initial Secret Sharing scheme for the cooperating participating. This operation is performed in the absence of the sender.

**Disenrollment:** Suppose the weight of an each player is decreased by one, then participants are combined to work together and upgrade all shares except in one share of the non-cooperative player. Therefore all shares are updated to be on a new secret function in a polynomial $\hat{f}(x)$, that share remains on the old Secret Sharing polynomial f(x), as a result, the share will not be valid.

**C. Trust Function:**

In the electronic commerce, major challenge is trust, "to believe a trusty" between a two or many different parties and how is their relationship between in global vision. "Trust" is a personal expectation an agent has about another's future behavior and its individual quality calculated based on the two agents concerned in the present and future. "Reputation" is a perception, which an agent had on other intentions [9].

The authors [9] had proposed, "A Trust party should increase their threshold α and keep cooperation increasing the reward to the maximum, when it decreases threshold β remains increasing the cost of defection. Reward and cost transaction fixed between α and β."

**Definition 2:** *Let $T_i^j$ (p) be the trust value assigned by $P_j$ to $P_i$ in period p. Let $T_i$: N → R be the trust function representing the reputation of $P_i$*

$$\mathbf{Ti(p)} = \frac{\mathbf{1}}{\boldsymbol{n-1}}\sum_{j\neq i} \boldsymbol{T}_i^j\,(\boldsymbol{p})$$

*Where $-1 \leq T_i(p) \leq +1$ and $T_i(0) = 0$. That is, we calculate the average of the trust values (personal quantity) in order to compute a player's reputation [5]*

Let's consider an example [5] to use the six possible keywords in the trust values of $P_1$, $P_2$, $P_3$ with respect to $P_4$ be $T_4^1$ (p) = 0.4, $T_4^2$ (p) = 0.5, $T_4^3$(p) = 0.6 .consequently , reputation of the trust value $P_4$ will be $T_4(p) = 0.5$. Public value $T_i(p)$ is assigned to each player $P_i$ that represents his reputation, i.e., $T_i(p) = T_i^j$ (p) for all j. Therefore they are three types of players that is B: bad, N: new and G: good with these six possible outcomes [9] values are defines a below table where α and β determine boundaries on the trust values for a different set of players, they can be approach then applies functions μ(x) and μ' (x) accordingly to update reputation of each player $P_i$, Parameters η, θ, and κ are used to increment or decrement the trust values. In intervals $[1-\varepsilon, +1]$ and $[-1, \varepsilon-1]$, functions μ(x)and μ' (x)both converge to 0 due to our assumption .The trust function is not just a function in the single round but the history it states that more the better a participants for example where $Ti(p) \in [\alpha, 1-\varepsilon]$, and penalize more than worse a participant is, e.g., Figure 2 : Defection, where $Ti(p) \in [\varepsilon-1, \beta]$. In addition, it provides opportunities for newcomers in the trust function but we do not know much about their behaviors and tragedies where $T_i(p) \in [\beta, \alpha]$.

| **Trust value** | **Cooperation** | **Defection** |
|---|---|---|
| $P_i \in B$ *if* $T_i(p) \in [-1, \beta)$ | Encourage | Penalize |
| $P_i \in N$ *if* $T_i(p) \in [\beta, \alpha]$ | Give a chance | Take a chance |
| $P_i \in G$ *if* $T_i(p) \in (\alpha, +1]$ | Reward | Discourage |

Table 1: Six Possible Action Functions in Trust [5].

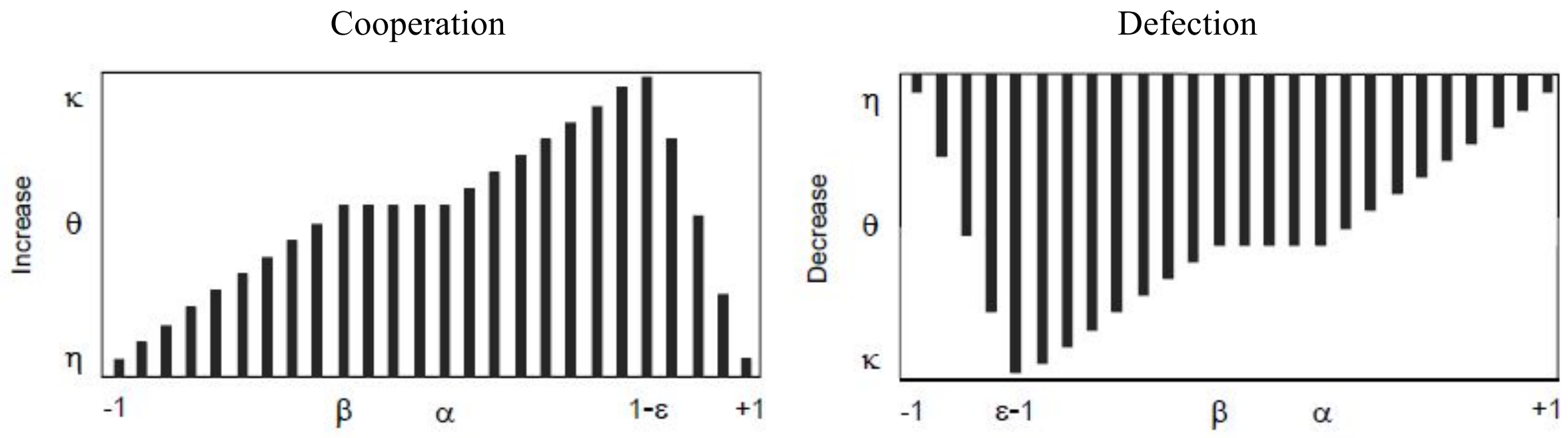

Trust value

Fig 2. Trust Adjustment by μ(x) and μ' (x) Functions [5].

Let $l_i \in \{0, 1\}$ where $l_i = 1$ denotes that player $P_i$ has cooperates with the current value and $l_i = 0$ denotes that he has defected. The proposed trust function [9] is as follows,

Where x = $T_i$ (p-1) 'x' is previous trust value.

$l_i = 1 \quad T_i(p) = T_i(p-1) + \mu(x)$; where

$$\mu(x) = \begin{cases} \theta - \eta/\beta + 1(x+1) + \eta & P_i \in B \\ \theta & p_i \in N \\ \kappa - \theta/\,1 - \in - \alpha \;\; (x-\alpha) + \theta & P_i \in G,\ T_i(p) \leq 1 - \in \\ \kappa/\in(1 - x - \in) + \kappa & T_i(p) > 1 - \in \end{cases}$$

$l_i = 0 \quad T_i(p) = T_i(p-1) - \mu(x)$

$$\mu^{|}(x) = \begin{cases} \kappa/\in(x+1) + \kappa & T_i(p) < 1 - \in \\ \theta - k/\beta - \in + 1 \;\; (x - € + 1) + k & P_i \in B,\ T_i(p) \leq 1 - \in \\ \theta & p_i \in N \\ \eta - \theta/1 - \alpha(x-\alpha) + \theta & P_i \in G \end{cases}$$

And for every function μ(x) and μ' (x)consists of four linear equations which are simply determined by two points $(x_1, y_1)$ and $(x_2, y_2)$

$$y = y_2 - y_1/x_2 - x_1 \; \frac{y2-y1}{x2-x1}(x - x_1) + y_1$$

From fig 2 factors like "reward" and "penalize" are greater than "encouragement" and "discouragement" by assigning $\eta = 0.01 < \theta = 0.05 < \kappa = 0.09$ defined at various points and trust function is calculated via regression .The authors [9] also determine parameters as the transaction cost to deal with cheap cooperation and expensive defections. For example consider a scenario in which every player co-operating a regular transaction for many times in order to get huge profits in the trust value. The authors can then defect in a critical transaction to severely damage the scheme. Therefore to adjust the trust value we need to define a transaction cost parameter, a weight for cooperation or defection [5].

## 7. Introduction to Cloud Computing

John McCarthy has given the basic general idea of the technology in 1960s, and wrote "computation may someday be organized as a public utility." In latter, grid computing concept is evaluated by the idea for making computer power as easy to access as an electric power grid. In 1990s cloud concept is introduced from telecommunications companies who made a radical shift from point to point data circuits to VPN (virtual private network) services. Prof Ramnath chellapa [11] defined the "Computing Paradigm where the boundaries of computing will be determined by economic rationale rather than technical limits alone". This is the basic idea what we refer till today whenever we discuss about cloud computing concept.

Cloud Computing [26, 27] is the use of computing resources of hardware and software that are delivered as a service over a network by typically the internet. The name comes from the use of a cloud shaped symbol as an abstraction for the complex infrastructure it contains in system diagrams. It also entrusts remote services with a user's data software and computation.

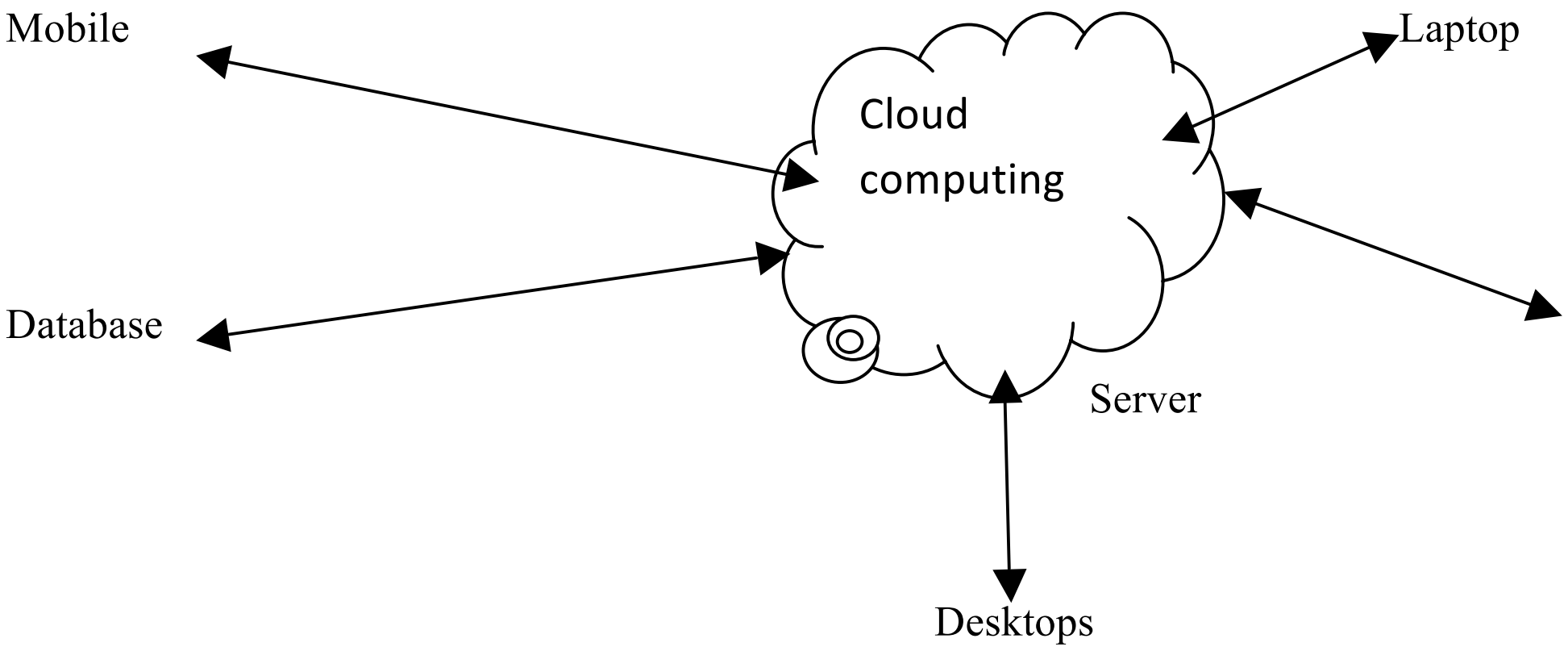

Fig 3 Cloud Architecture [26, 27]

Cloud computing relies on sharing of resources to acquire the coherence and economies of scale over a network as in the electricity grid computing and it is a wide concept to gather all infrastructure and shared services.

## 7.1 Cloud computing in cryptography

Cloud computing is widely used in the IT business and many more workplaces, in that, cases file sharing , file transferring is done among different groups of people in the network, therefore these have a very highly confidently of security to do a secured task. For example, IDC's 2008 cloud services user survey [12], executives had cited a number of different security challenges among the cloud users. Cloud Security Alliance's [13] initial report contains all different sort of cytology based on 15 different security domains and processes that need to be followed in an overall cloud deployment. Security concerns are

- Traditional security
- Availability
- Third-party Data Control

At the same time there are also some security issues that can be encounter. Such as, we cannot depend on a single security in the cloud computing though traditional encryption can solve privacy-related problems that come into the new context. So that we will need to provide more secured functional cryptographic techniques. There are some cryptographic techniques that play a vital role in cloud computing security such as searchable encryption, proxy decryption, attribute-based encryption, Holomorphic encryption, and many others.

## 8. Application in Cloud Computing

From the NIST [28] the definition of cloud computing defines three service models. They are

1. Cloud infrastructure as a service (IaaS): Infrastructure as a Service provides various infrastructures which contain hardware, storage and other computing resources. It provides to the consumer for processing, storage where, consumer is able to deploy and run selected software. They do not manage the hosted firewalls in the networking components.
2. Cloud platform as a service (PaaS): Platform as a Service provides a ready to use platform these includes the all operation systems which consumers provides to the infrastructure and they also built applications on development frameworks and some functional databases. Consumer does not manage or control the underlying cloud infrastructure.
3. Cloud software as a service (SaaS): Software as a Service provider provides an operating environment used to deliver a complete application in web- based applications. Apps related are email consumer does not manage or control any networks, servers, storage or OS. They can just use possible graphical user interfaces [14].

### 8.1 Types of file systems in cloud computing

**Google File System**

The Recovery on the distributed file system developed by Google and specially designed to provide useful and reliable access to data, using large server clusters. The GFS architecture consists of three elements: Clients, Masters and chunk servers. It consists of a single master and multiple chunk servers that are remote by multiple clients [15].

**Amazon S3**

The Amazon Simple Storage Service is a distributed storage systems based on Dynamo [29]. The Dynamo uses key value storage in a Distributed Hash Table (DHT) and has no support association schemes. We can make a secure level of scalability and availability, data is divided and duplicated into multiple machines, using a consistent hashing, being consistency facilitated by multiple versions of objects [15].

**Microsoft Azure**

Microsoft SQL Azure is compound of a set of services for data storing and processing data in cloud. The storage devices can access and balance of the load is done automatically through a set of nodes responsible for physical storage, providing scalability and availability. Consumer needs to create a storage account, which can be obtained from the Windows Azure portal web interface [15].

By using the “computing”, “storage” and “software” as a service the consumers are not required to invest in IT infrastructure on their own. A serious challenge in cloud computing is “resource management”. The consumers are expecting in terms of resource availability from the every transactions they perform. Platforms like vimeo, flickr,

Slideshare, skype, etc., are listed as cloud applications; most of these platforms share the data (file transferring, image sharing, video, voice). However they are many more apps that might be less aware of it. They are some enterprises that provide money valuations in the cloud and satisfy their demands. The services are provided and try to maximize their own costs profits in situation of the data is prioritizing the consumers jobs. These factors may leads to competition, negotiation, dynamic allocation, and automatic load balancing [5].

In the cloud the secure system using threshold Secret Sharing is done in the new method therefore, there exist some questions like how the systems can be automated that configures the availability of dissimilar components. Thus can be help to better comply with the service level agreements (SLA), it is a software serves as the foundation between cloud providers and consumers.

There are some challenges in the cooperative game between the cloud providers and consumers, which are classified into:

1) The set rules are followed by the service providers in service- level agreements.
2) Consumers receive their services with a high satisfaction rate.

Here is a scenario from cloud computing that influence the computer gaming industry. These gaming industries mostly are the online games and they do very huge business for cost profits the flexibility [16].Lets consider a 'p' is new creator in online multi-player games, wants to utilizes a computing cloud to deploy the core gaming for their previous game. Now 'p' doesn't have any idea about the game whether public are going to accept it or not. Therefore 'p' has committed to choose a cloud computing platform for automatic scaling of a game. Eventually 'p' is ready to get a decent response time but the cost maximum for the threshold value at hourly in order to maintain their limited constraints. If response time is not met to the constraints likely 'p' loses some gamers, hence it will penalize the cloud provider in case of a violation. 'r' provides cost services and quality measurements services for the resource usage calculation of consumers. Now 'p' will handle the all financial cost calculations to 'q', it is hired by 'p' as trusted third party for resource usage. This typical scenario requires that 'p', with their cloud provider 'r', create a SLA

- The Max hourly cost ($C_{max}$) need to below the $Thr_{max}$
- Average response time ($RT_{avg}$) need to below $Thr_{max}$ with respected to above
- Max response time ($RT_{max}$) need to below $Thr_{max}$

Thus 'r' is not capable of calculating the composite metrics in SLA .However it is capable of providing the running time and unit costs during negotiation process. Therefore 'r' is defined as the third party to look after the 'p' cost services. $C_{max}$ calculates the data from cost services and usage services. In case of SLA violations, management service is noticed and 'q' accounting is contacted for the financial penalties [15].

We can refer to excessive spike in online shopping with "Amazon" at the end of the year. It would be easier and flexible for the both consumers and service providers if the system takes can play the automatic configuration strategy and relies less on busier components during certain periods and limits in the configuration. Here comes a question, how this can be done by the continuous interactions between the providers and consumers.

In the distributed secure system using the threshold secret sharing even if the some servers do not act properly due to an adversarial attack or delay from the response time in the system, it can still perform the task if the certain number of components operates appropriately. Therefore, we intend to show this cooperation can be changed in the secret sharing scheme. In other words, the consumers use a reputation management from the system to rate different components in the systems. Further, the system is reconfigured over the cloud to have a guarantee the service-level agreement [5].

The author [5] consists of a dealer who initiates a weighed secret sharing scheme, *n* cloud providers denoted by $P_1, \ldots, P_n$ and many servers interacting with the cloud providers. Let $r = (r_1, r_2, \ldots, r_n)$ and $w = (w_1, w_2, \ldots, w_n)$ be the vector of players' trust values and the vector of players' weights accordingly. The initial values in '*r'* are going to be zero (i.e., all service providers are treated as newcomers), whereas the initial values in '*w'* are chosen by the dealer based on a specific distribution. We first define the following actions where each player's action $A_i \in \{C, D, X\}$:

1) *C*: for *cooperative* players where $P_i$ is available at the required time and he sends correct shares to other parties.
2) *D*: for *uncooperative* players where $P_i$ is not available at the required time or he responds with delay.
3) *X*: for *corrupt* players where $P_i$ has been compromised by an adversary and he may send incorrect shares.

Let assume a secret key $\xi$ is selected in order to accomplish a secure task whenever it is required. For instance, we can refer to secure auctions in which bidders submit their sealed-bids to auctioneers when the auction starts and then the auctioneers define the outcomes (i.e., the winner and the selling price) without revealing the losing bids.

We can therefore assume that secret key $\xi$ is used by many auctioneers to start or accomplish several sealed-bid auctions overtime on behalf of a seller. Considering this secure auction scenario, a dealer (or a seller) first distributes shares of this secret among different service providers (or clouds) according to their initial weights in vector $w_3$, as shown in Figure 3. The author [5] then leaves the scheme.

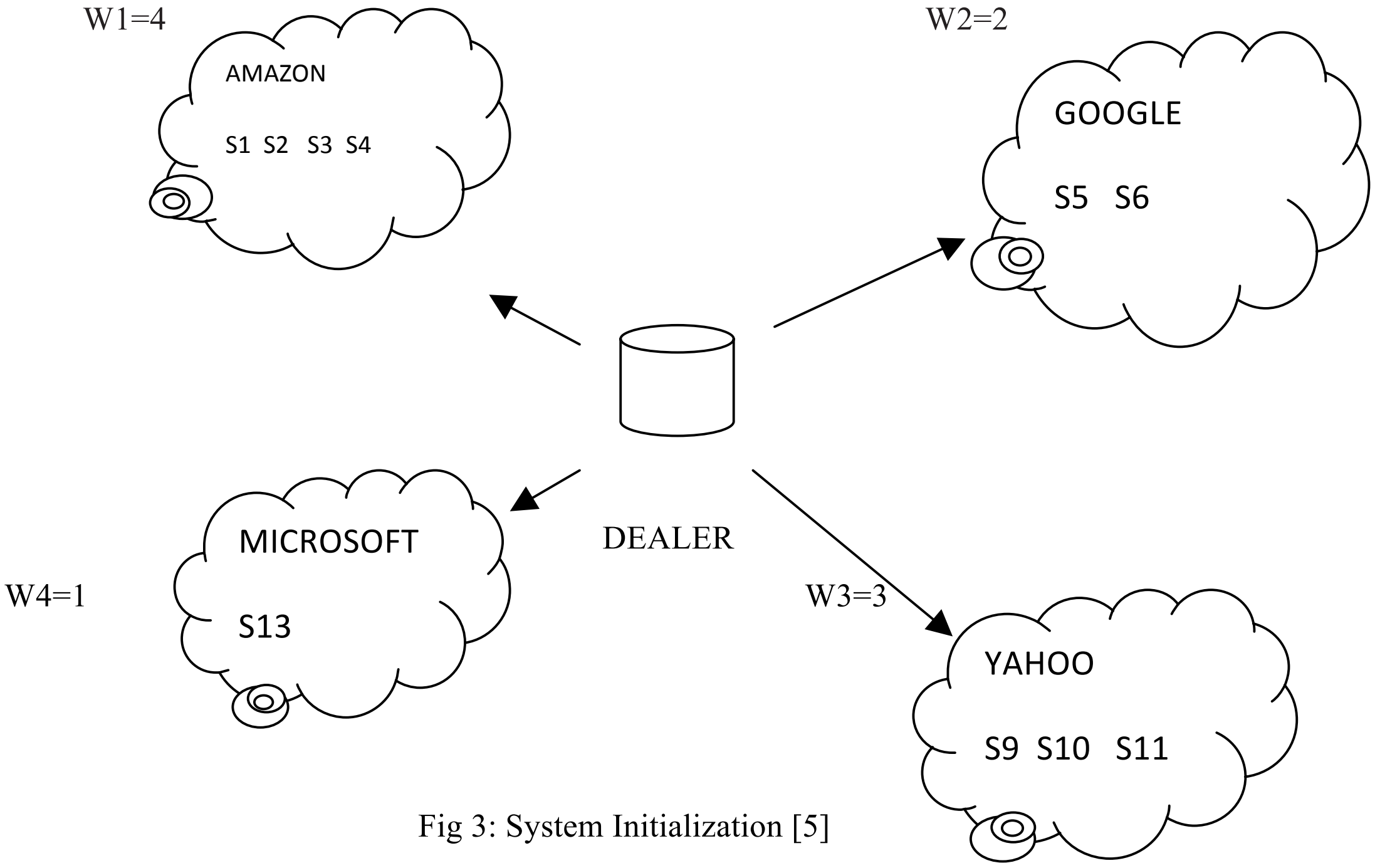

Fig 3: System Initialization [5]

From the above system initialization, different servers or auctioneers interact with the cloud providers to perform their tasks in the absence of dealer from time to time. Requests for these shares are sent to the cloud providers by the server. The secret is recovered on these servers and then a secure procedure or sealed-bid auction is accomplished and finally the secret and its corresponding shares are erased from the servers and keep servers busy at the all-time check care of not go into idle state.

Trust – to- share ratio computation [9] this values may affects the weight of a players where actions $A_i \in \{C, D\}$ as well as a trust function and these servers rate each component of the cloud in terms of its response time .The main issues is going to be more critical in real time systems where response time plays an important role. In some case of corruption $A_i = X$ values will be rebooted. Then values return to the scheme and are treated as newcomers as we illustrated earlier. Corrupted actions therefore sending incorrect shares are detectable by using a Verifiable Secret Sharing Scheme [7].

In the last phase, the service providers jointly collaborate to reconfigure the scheme from new weights and they initially enroll the new shares by using a protocol (enrollment) share of share $S_{14}$ is enrolled for the fourth party. Subsequently, shares are updates except the shares that are scheduled to be disembroiled and they are transformed to a new Secret Sharing polynomial. If suppose same share $S_{14}$ is not updated then the first player is going to have three shares afterward. The main benefit and reason for using the threshold Secret Sharing in a distributed secure system is its fault tolerance and availability. If one component is compromised by an adversary or he responds with delay, other participants can carry out the intended procedure [5].

## 9. Trust Function

In Social Secret Sharing [10], some social characteristics and functions are in the new trusted function. Properties which are to be adjusted in order to adjust the trust values in different cases are [5]

1. *Type*: $\alpha$ and $\beta$ parameters are used to classify the players in three sets *B, N, G*. As a result, six scenarios are considered to increment or decrement the trust value.
2. *History*: $T_i(p)$ represents a history is action taken by a player $P_i$ in the trusted function so far, therefore the quality of a good player ranges from α all the way to +1 which characterize how good a players is and what kind of history the player has before the trust function applies an single evaluation strategy. It is intended to use a social evaluation strategy by adding the following properties to our new trust function.

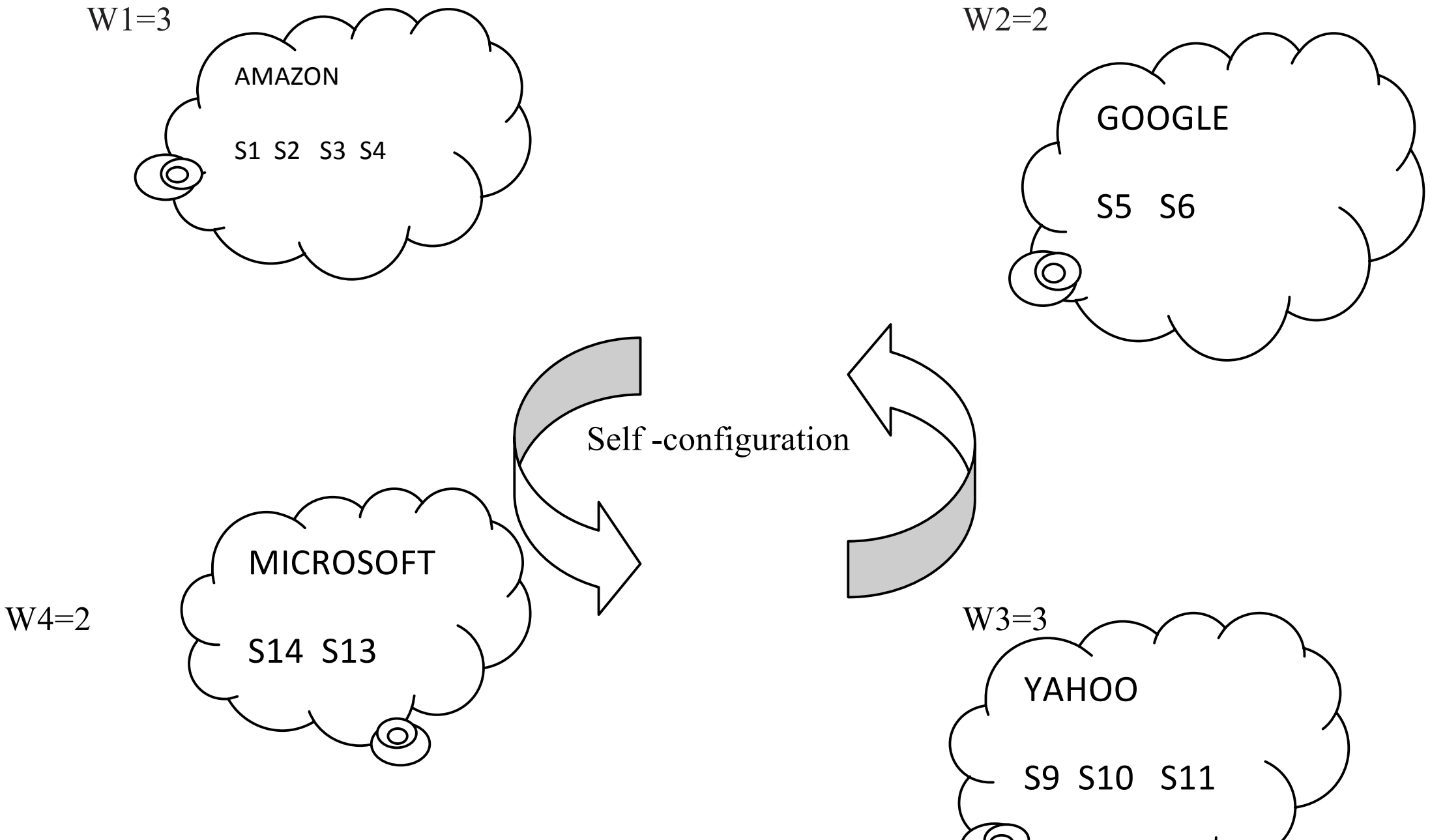

Fig 4: Self –Configuration [5]

3. Sociality: Social behaviors of the some other parties each time in the trusted function and it is used from words beside player's type and history this function.

   Consider all players together for trusted new function should be satisfying the following social conditions and those are [5]

- If $\delta = n$, i.e., all players have cooperated, it is not required to increase the trust value of anyone.
- If $\delta = 0$, i.e., all players have defected, it is not required to decrease the trust value of anyone.
- If $\delta > n/2$, i.e., majority of the players have cooperated, cooperation should be rewarded less and defection should be penalized more.
- If $\delta < n/2$, i.e., majority of the players have defected, defection should be penalized less and cooperation should be rewarded more.
- If $\delta = n/2$, i.e., the number of cooperative players and non-cooperative ones are equal, cooperation and defection should be rewarded and penalized with an equal ratio.

Social trust function is termed from the trusted function using the previous μ(x) and μ' (x) function which is explained in earlier in the trusted function

$$T_i(p) = \begin{cases} T_i(p-1) + (1 - \delta/n)\mu(x) & \text{if } l_i=1 \\ T_i(p-1) - (\delta/n)\mu'(x) & \text{if } l_i=0 \end{cases}$$

An example from table 2, At each time players wants to gains at regular intervals of their rewards(25%) that is proportional to the number of "non cooperative" players, if player lose then trust value is (75%) that is proportional to the number of "cooperative" players in the social trust function [5].

| $\delta = \sum^{n}_{i=1} l_i$ | **Cooperation** | **Defection** |
|---|---|---|
| N | $T_i(p-1)$ | No defection |
| 3/4n | $T_i(p-1)$+0.25 $\mu(x)$ | $T_i(p-1)$-0.75 $\mu'(x)$ |
| 1/2n | $T_i(p-1)$+0.5 $\mu(x)$ | $T_i(p-1)$-0.5 $\mu'(x)$ |
| 1/4n | $T_i(p-1)$+0.75 $\mu(x)$ | $T_i(p-1)$-0.25 $\mu'(x)$ |
| 0 | No cooperation | $T_i(p-1)$ |

Table 2: Computing $T_i$ (P) with Different values of δ [5]

Reputation is a social quantity which gives the all information about the player's history. Thus it is valid assumptions that cooperation has more value if majority of the players and defecting, similarly human social life is which cooperation is appreciated more when most players are not cooperating. Further if all players are defecting the trust value should remain unchanged. Eventually competition eliminations justified by the uniformity of the actions [5].

## 10. Conclusion

The Social properties will produce a new model for system management in the distubuted secure schemes and shows Social Secret Sharing in cloud computing with a new trust function [5].Various tools in the IT infrastructure which need to have a security during their any kind of transaction in the cloud computing are shown their problems with existing models. I also conclude that all related topics regarding the Secret Sharing Schemes have implemented a Secret Sharing using numbers, threshold values and number of shares to disturb the secret among group. As a result, we came to know, secret can be constructed with the trusted mathematically and shared among in the new trusted function and can be used to access and maintain confidently in cloud computing.

## 11. References

1 Kirill Morozov " Introduction to Secret Sharing" , Kyushu university and Lecturer notes and example by lidong Zhou http://www.cs.cornell.edu/courses/cs513/2000sp/SecretSharing.html.

2 R. J. McEliece and D. V. Sarwate. On sharing secrets and Reed-Solomon codes. Communications of ACM, 24(9):583–584, 1981.

3 Shamir, "How to share a secret," Communications of the ACM, vol. 22, no. 11, pp. 612–613, 1979.

4 H. Ghodosi, J. Pieprzyk, and R. Safavi-Naini. Remarks on the multiple assignment Secret Sharing scheme. In Y. Han, T. Okamoto, and S. Qing, editors, ICICS '97: First International Conference on Information and Communication Security, volume 1334 of Lecture Notes in Computer Science, pages 72–80. Springer-Verlag, 1997.

5 Nojoumian M. and Stinson D. R., Social Secret Sharing in Cloud Computing Using a New Trust Function, 10th IEEE Annual Conference on Privacy, Security and Trust (PST), pp. 161-167, Paris, France, 2012.

6 Nojoumian M., Stinson D. R., and Grainger M., Unconditionally Secure Social Secret Sharing Scheme, IET Information Security (IFS), Special Issue on Multi-Agent and Distributed Information Security, vol. 4, issue 4, pp. 202-211, 2010.

7 B. Chor, S. Goldwasser, S. Micali, and B. Awerbuch, "Verifiable Secret Sharing and achieving simultaneity in the presence of faults," in 26th Annual IEEE Symposium on Foundations of Computer Science FOCS, 1985, pp. 383–395. IEEE Computer Society, 1985.

8 Figure 1 "Secret Sharing and Cloud Computing" Workshop Overview Workshop Organizers Kyushu University, Institute of Mathematics for Industry Prof. Tsuyoshi Takagi, Assistant Prof. Kirill Morozov,June 7(Tue),2011

9 Nojoumian M. and Lethbridge T. C., A New Approach for the Trust Calculation in Social Networks, 3rd International Conference on E-Business (ICE-B), pp. 257-264, Setubal, Portugal, 2006.

10 Nojoumian M. and Stinson D. R., Brief Announcement: Secret Sharing based on the Social Behaviors of Players, 29th ACM Symposium on Principles of Distributed Computing (PODC), pp. 239-240, Zurich, Switzerland, 2010.

11 Dr. Ramnath Chellappa is currently an Associate Professor and Caldwell Research Fellow in the Information Systems & Operations Management area at the Goizueta Business School, Emory University, 1990.

12 IT Cloud Services User Survey, pt.2: Top Benefits & Challenges. http://blogs.idc.com/ie/?p=210 by IDC survey, august 2008.

13 CSA (cloud Security alliance) Security Guidance for Critical Areas of Focus in Cloud Computing. http://www.cloudsecurityalliance.org/guidance/csaguide.pdf, 2011

14 http://www.apptis.com/documents/An%20Intro%20to%20Cloud%20Computing%20in%20the%20Public%20Sector-3-14-11.pdf by Apptis, 2010-2011.

15 Edna Dias Canedo, Rafael Timoteo de Sousa Junior and Robson de Oliverira Albuquerque, "Trust Model For Reliable file exchange in cloud computing," Electrical Engineering Department, University of Brasília UNB – Campus Darcy RibeiroAsa Norte Brasília DF, Brazil, 70910-900,Feb 2012.

16 Kennedy, S.: Denis dyack's head is in the clouds [http://tinyurl.com/n2gg2w](2009)

17 Keith M. Martin , "Challenging the adversary model in Secret Sharing Schemes" from information security group, Department of Mathematics, Royal Holloway, University of London , Egham surrey TW20 0EX ,United Kingdom,2008.

18 G. Blakley, "Safeguarding Cryptographic Keys", in *Proc. American Federation of Information Processing Societies* (*AFIPS*) Arlington, VA, June 1979, vol. 48, pp. 313-317.

19 C. Ding, D. Pei, and A. Salomaa, *Chinese Remainder Theorem: Applications in Computing, Coding, Cryptography,* Singapore: World Scientific Pub Co., 1999.

20 J. Benaloh, "Secret Sharing homomorphism's: Keeping Shares of a Secret", *Advances in Cryptology-CRYPTO'86*, vol. 263, pp. 251-260, 1986.

21 M. Naor, and A. Shamir, "Visual Cryptography", *Advances in Cryptology-EUROCRYPT'94*, pp. 1-12, 1994.

22 SY. Desmedt and Y. Frankel. Threshold cryptosystem. In G. Brassard, editor, *Advances in Cryptography—Crypto'89*, volume 435 of *Lecture Notes in Computer Science*, pages 307–315. Springer-Verlag, 1990.

23 D. R. Stinson and S. A. Vanstone, A Combinatorial Approach to Threshold Schemes, SIAM J. Disc.Math., 1(1988), 230-236.

24 A. Beimel, and B. Chor, "Universally Ideal Secret Sharing Schemes", *IEEE Transactions on Information Theory,* vol. 40, pp. 786-794, 1994.

25 A. Herzberg, S. Jarecki, H. Krawczyk, and M. Yung, "Proactive Secret Sharing or: How to cope with perpetual leakage," in *15th Annual International Cryptology Conference CRYPTO*, ser. LNCS, vol. 963.Springer, 1995, pp. 339–352.

26 M. Armbrust, A. Fox, R. Griffith, A. Joseph, R. Katz, A. Konwinski, G. Lee, D. Patterson, A. Rabkin, I. Stoica, M. Zaharia. Above the Clouds: "A Berkeley View of Cloud computing". Technical Report No. UCB/EECS-2009-28, University of California at Berkley, USA, Feb. 10, 2009.

27 L. Youseff, M. Butrico, and D. Da Silva. "Toward a Unified Ontology of Cloud Computing," Grid Computing Environments Workshop (GCE '08), pp. 1—10 (2008).

28 P. Mell and T. Grance. "The NIST Definition of Cloud Computing". National Institute of Standards and Technology (2009).

29 DeCandia, G., Hastorun, D., Jampani, M.,Kakulapati, G.,Lakshman, A.,Pilchin, A.,Sivasubramanian, S., Vosshall, P., and Vogels. "Dynamo: amazon's highly available key-value store". Proceedings of twenty-first ACM SIGOPS symposium on Operating systems principles. ACM. New York, NY, USA. 2007.

30 T. Rabin and M. Ben-Or, "Verifiable secret sharing and multiparty protocols with honest majority," in *21st Annual ACM Symposium on Theory of Computing STOC*, 1989, pp. 73–85.

31 D. Beaver, "Multiparty protocols tolerating half faulty processors," in *9th Annual International Cryptology Conference CRYPTO*, ser. LNCS, vol. 435. Springer, 1989, pp. 560–572.

32 J. C. Benaloh and J. Leichter, "Generalized secret sharing and monotone functions," in *8th Annual International Cryptology Conference CRYPTO*, ser. LNCS, vol. 403. Springer, 1988, pp. 27–35.